\documentclass{goeproc}
\usepackage{natbib}
\bibpunct{(}{)}{;}{goeproc}{}{,}
\usepackage{shortvrb}
\MakeShortVerb{\|}
\def\pdfLaTeX{pdf\kern.06em\LaTeX}
\begin{document}

\title{Ray tracing of ion-cyclotron waves in a coronal funnel}

\author[*]{R.~Mecheri}
\author[ ]{E.~Marsch}

\affil[ ]{Max-Planck-Institut f\"ur Sonnensystemforschung,
Katlenburg-Lindau, Germany}

\affil[*]{\textit{Email:} mecheri@mps.mpg.de}

\runningtitle{Ray tracing of ion-cyclotron waves in a coronal
funnel} \runningauthor{R.~Mecheri and E.~Marsch}

\firstpage{1}

\maketitle

\begin{abstract}
Remote observations of coronal holes have strongly implicated the
kinetic interactions of ion-cyclotron waves with ions as a principal
mechanism for plasma heating and acceleration of the fast solar
wind. In order to study these waves, a linear perturbation analysis
is used in the work frame of the collisionless multi-fluid model. We
consider a non-uniform background plasma describing a funnel region
and use the ray tracing equations to compute the ray path of the
waves as well as the spatial variation of their properties.
%
%
\end{abstract}

\section{Introduction}
The ultraviolet spectroscopic observations made by SUMER and UVCS
aboard SOHO indicated that heavy ions in the coronal holes are very
hot with high temperature anisotropy \cite[see,
e.g.,][]{mecheri-goe:Kohl97, mecheri-goe:Wilhelm98}. This result is
a strong indication for heating by ion-cyclotron resonance
\citep[i.e., collisionless energy exchange between ions and wave
fluctuations, see][Chap~10]{mecheri-goe:Stix} involving
ion-cyclotron waves
that are presumably generated in the lower corona from small-scale
reconnection events \citep{mecheri-goe:Axford95}. Performing a
Fourier plane wave analysis, ion-cyclotron waves are studied using
the collisionless multi-fluid model. While neglecting the electron
inertia, this model permits the consideration of ion-cyclotron wave
effects that are absent from the one-fluid MHD model. Realistic
models of density and temperature as well as a 2D funnel model
describing the open magnetic field are used to define the background
plasma. Considering the WKB approximation, we first solve locally
the dispersion relation and then perform a non-local wave analysis
using the ray-tracing theory, which allows to compute the ray path
of the waves in the funnel as well as the spatial variation of their
properties.

\section{Basic equations}

\begin{figure}
\begin{center}
$\begin{array} {c@{\hspace{-0.1in}}c} \hspace{-1.cm}
\includegraphics[width=7.cm]{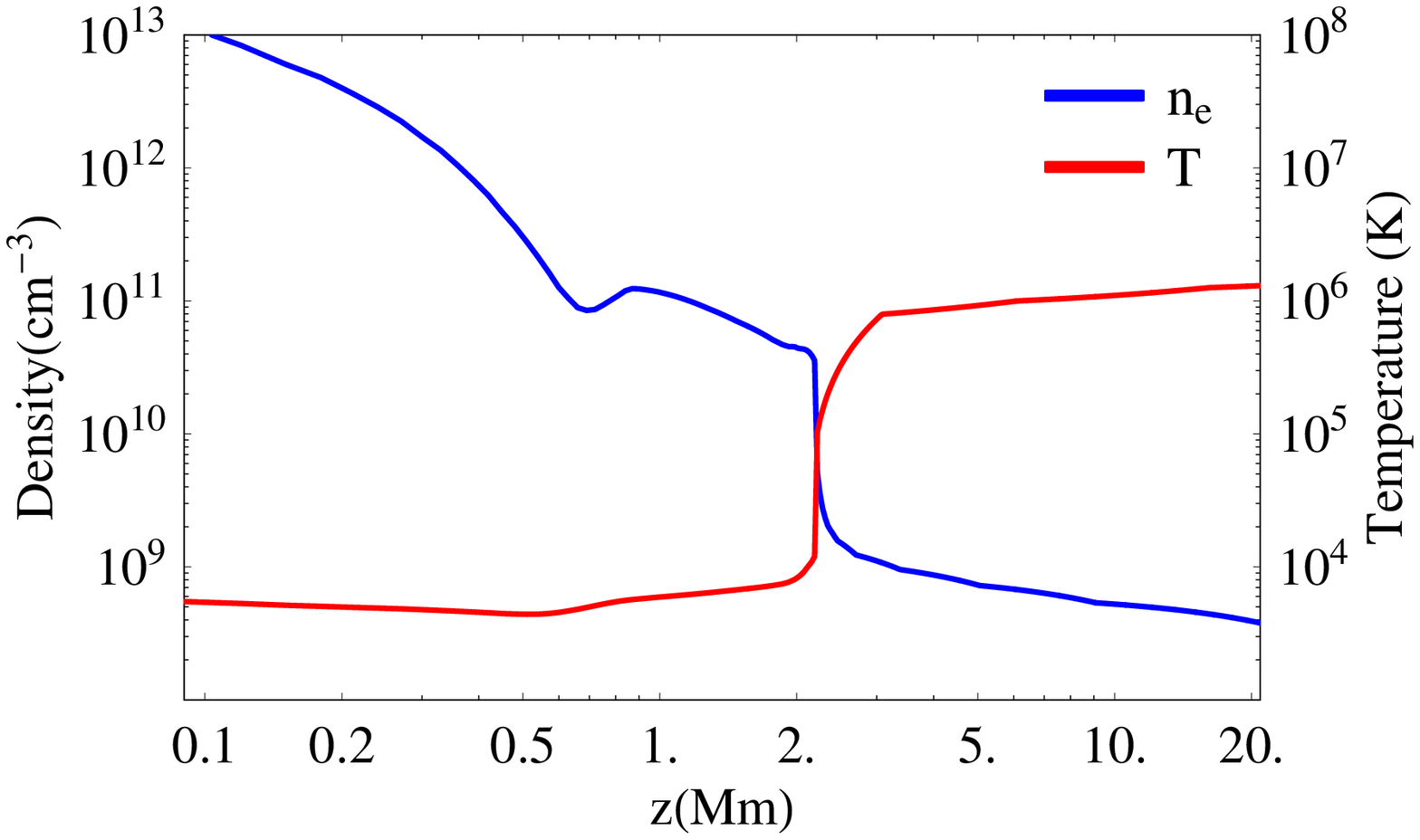}
\end{array}$
\hspace{-0.2cm} $\begin{array} {c@{\hspace{-0.1in}}c}
\includegraphics[width=5.2cm]{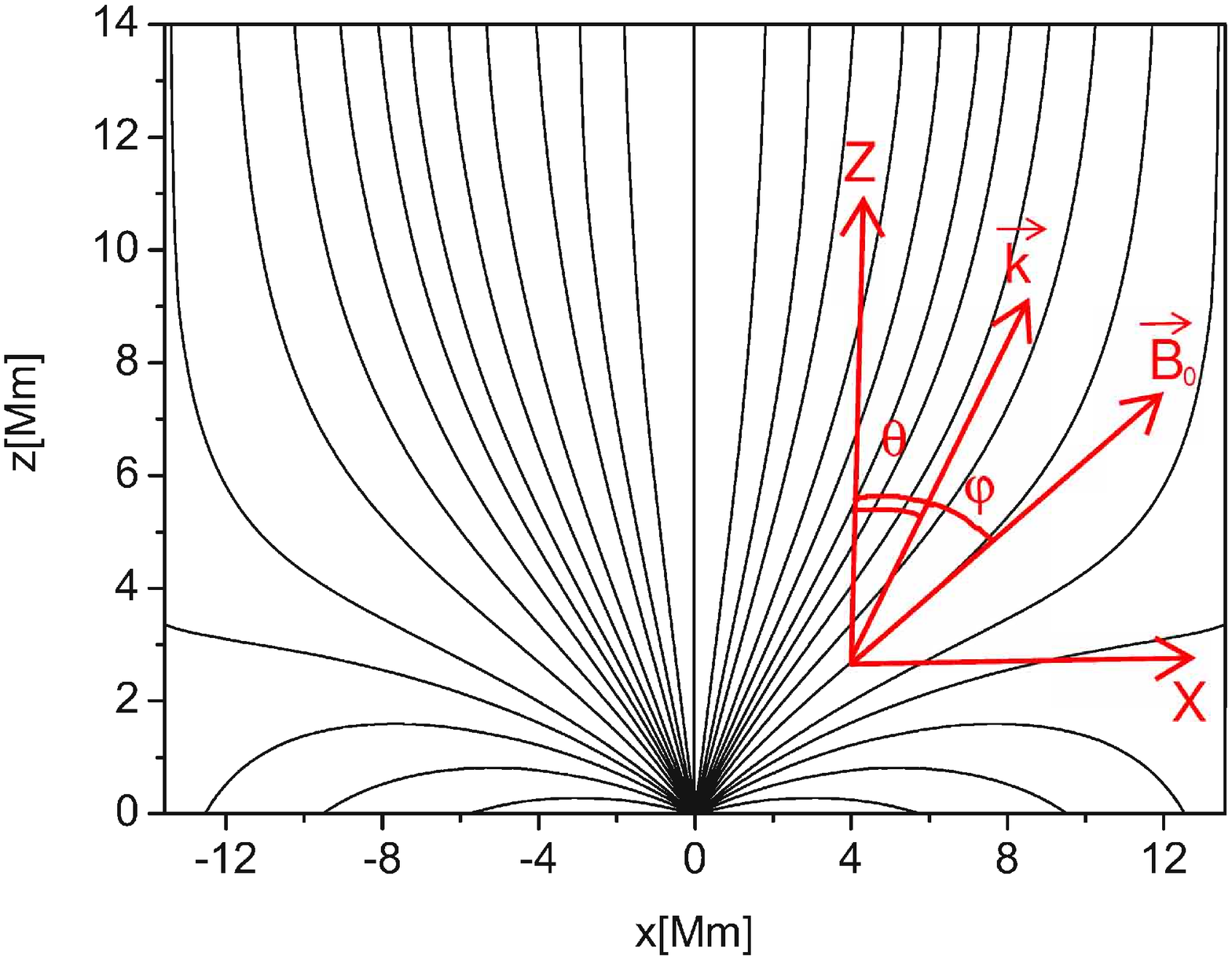}
\end{array}$
\end{center}
\vspace{-0.5cm} \caption{\small Left: Electronic density ($n_{e}$)
and temperature (\textit{T}) model-profiles of the chromosphere
\citep{mecheri-goe:Fontenla} and the lower corona
\citep{mecheri-goe:Gabriel}. Right: Funnel magnetic field geometry
as obtained from a 2-D potential field model
\citep{mecheri-goe:Hackenberg}. 
}
\label{funnel}
\end{figure}

The cold collisionless fluid equations for a particle species $j$
are:
\begin{equation}
  \frac{\partial n_{j}}{\partial
  t}+\nabla\cdot(n_{j}\textbf{v}_{j})=0\,,
\end{equation}
\vspace{-0.3cm}
\begin{equation}
  m_{j}n_{j}(\frac{\partial \textbf{v}_{j}}{\partial
  t}+\textbf{v}_{j}\cdot\nabla \textbf{v}_{j})-q_{j}n_{j}
  (\textbf{E}+\textbf{v}_{j}\times
  \textbf{B})=0\,,
\end{equation}\\[-0.3cm]
where $m_{j}$, $n_{j}$ and $\textbf{v}_{j}$ are respectively the
mass, density and velocity of a species $j$.
The electric field $\textbf{E}$ and the magnetic field $\textbf{B}$
are given by Faraday's law, i.e. $\nabla\times
\textbf{E}=-\partial\textbf{B}/\partial t$. The background density,
temperature and magnetic field are described in Fig.~\ref{funnel}.
Considering the quasi-neutrality and no ambient electric field, we
perform a Fourier plane-wave analysis. The dispersion
relation is obtained using
\begin{equation}
D(\omega,\textbf{k}
,\textbf{r})=Det[\frac{c^{2}}{\omega^{2}}\textbf{k}
\times(\textbf{k}\times\textbf{E})+
\vec{\varepsilon}(\omega,\textbf{k},\textbf{r})\cdot\textbf{E}]=0\,,
\label{dispesion}
\end{equation}
where $c$ is the speed of light in vacuum, $\vec{\varepsilon}$ is
the dielectric tensor which is a function of the wave frequency
$\omega$, the wave vector $\textbf{k}$ and the large-scale position
vector $\textbf{r}$. The wave vector $\textbf{k}$ lies in the $x-z$
plane, with $\textbf{k}=k(sin\theta,0,cos\theta)$.

The dispersion relation (\ref{dispesion}) is used to compute the
Hamiltonian-type ray-tracing equations \citep{mecheri-goe:Weinberg}:

\begin{equation}
\frac{\textrm{d}\textbf{r}}{\textrm{d}t}=-\frac{\partial
D(\omega,\textbf{k} ,\textbf{r})/\partial \textbf{k}}{\partial
D(\omega,\textbf{k} ,\textbf{r})/\partial \omega},~
\frac{\textrm{d}\textbf{k}}{\textrm{d}t}=\frac{\partial
D(\omega,\textbf{k} ,\textbf{r})/\partial \textbf{r}}{\partial
D(\omega,\textbf{k} ,\textbf{r})/\partial \omega}~\cdot
\end{equation}

\section{Numerical results}\label{results}
\begin{figure}
\begin{center}
$\begin{array} {c@{\hspace{-0.35in}}c}
\hspace{-0.4cm}\includegraphics[width=5.cm]{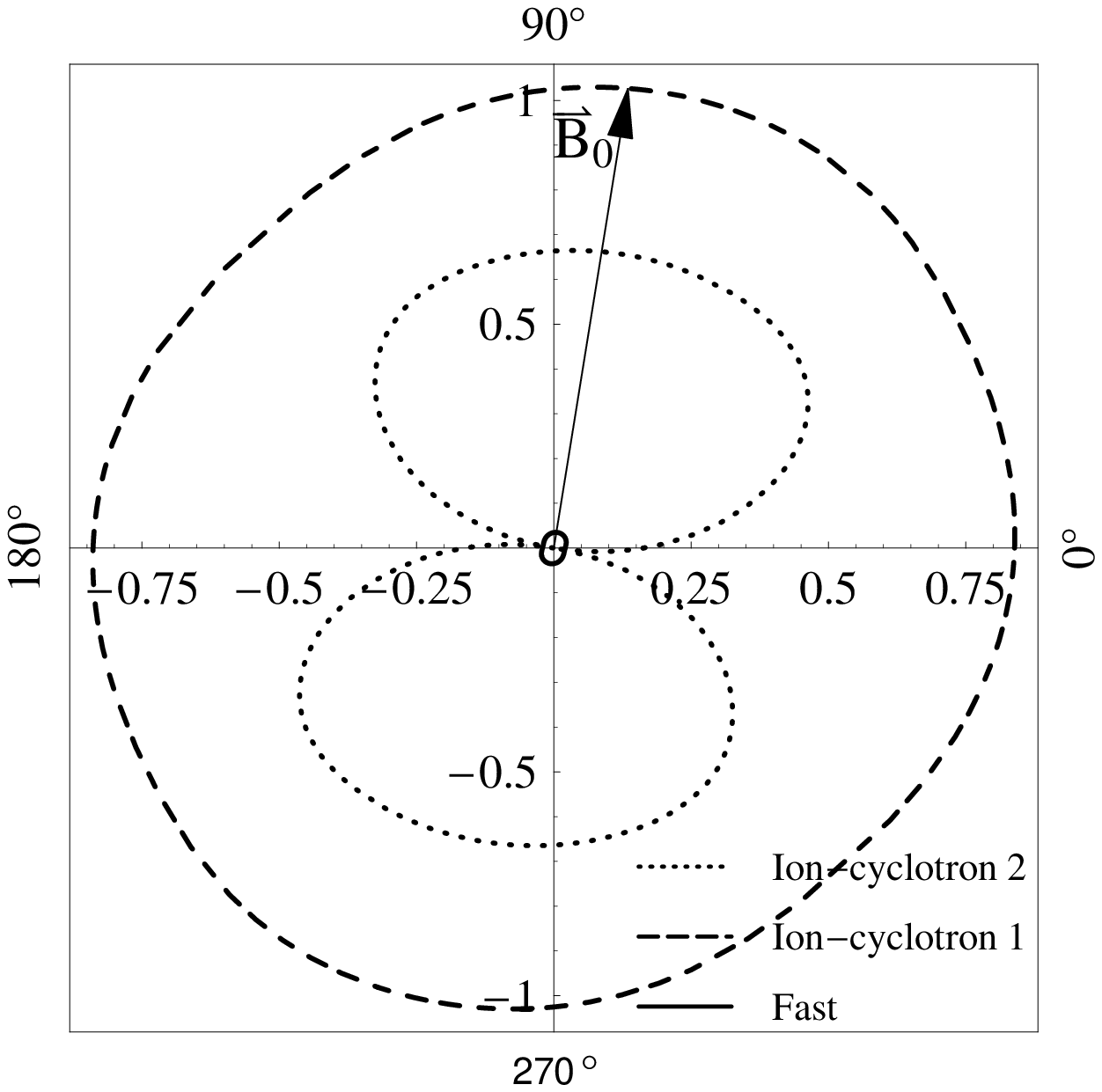}
\end{array}$
\hspace{0.5cm} $\begin{array} {c@{\hspace{-0.1in}}c}
\includegraphics[width=7.2cm]{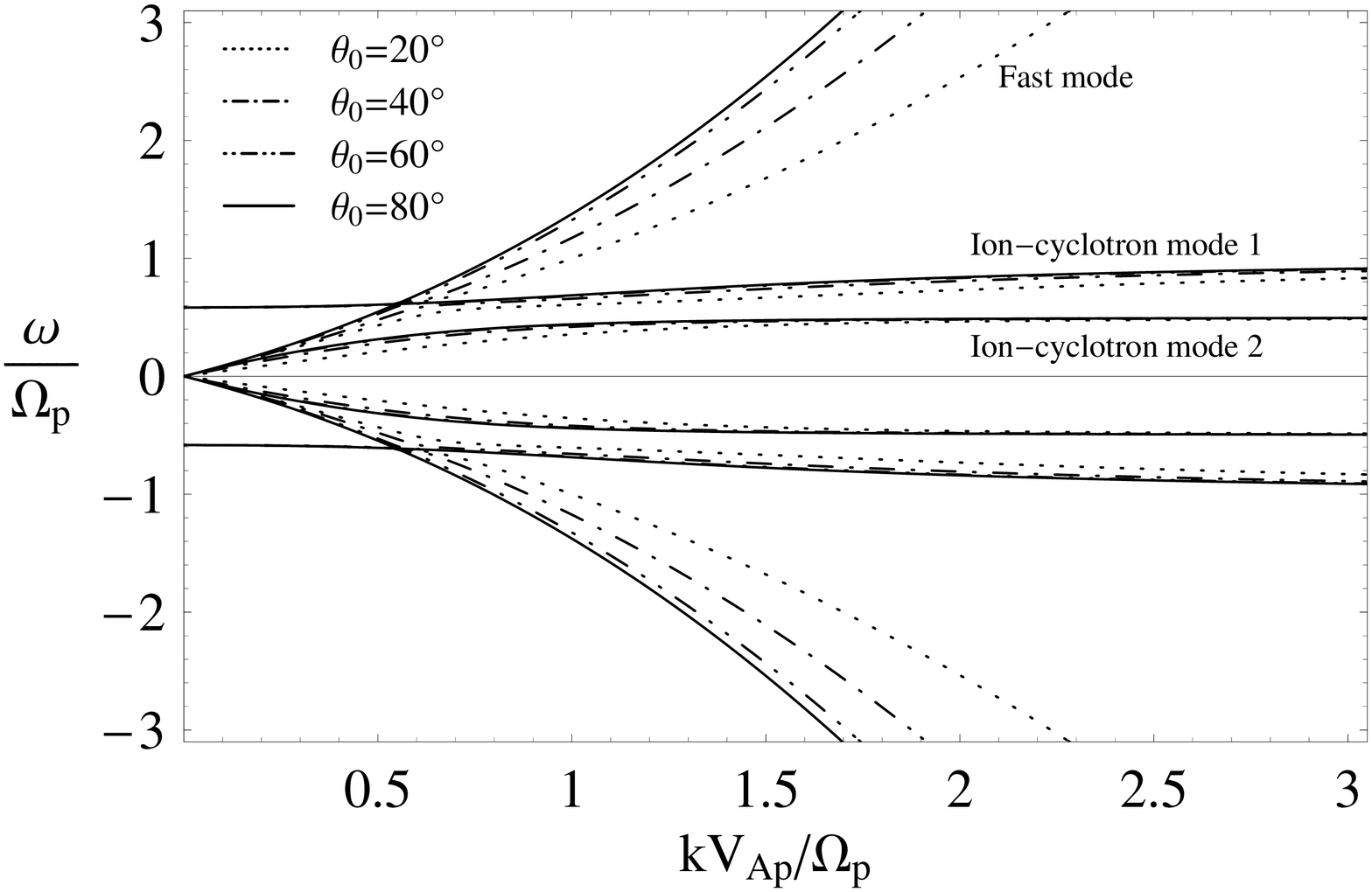}
\end{array}$
\end{center}
\vspace{-0.8cm} \caption{\small Right: Dispersion curves in the cold
three-fluid (e-p-He$^{2+}$) model at the location (x=7.5 Mm, z=2.5
Mm) in the funnel with a \textbf{B}$_{0}$-inclination angle
$\varphi\approx79^{\circ}$ and for different propagation angles
$\theta$ ($\Omega_{p}$ and V$_{Ap}$ are the proton cyclotron
frequency and Alfv\'{e}n speed). Left: Polar plot of the group
velocity as a function of $\theta$ and for
$kV_{Ap}/\Omega_{p}=0.2$.} \label{VGR}
\end{figure}
\begin{figure}
\begin{center}
$\begin{array} {c@{\hspace{-0.01in}}c@{\hspace{-0.01in}}c}
\multicolumn{1}{l}{~~~~~~~\hspace{-1.5cm}\mbox{Ion-cyclotron mode
2}} & \multicolumn{1}{l}{\hspace{+1.2cm}\mbox{Ion-cyclotron mode 1}}
&
\multicolumn{1}{l}{\hspace{+1.2cm}\mbox{Fast mode}}\\[-0.2cm]
\end{array}$
$\begin{array} {c@{\hspace{-0.1in}}c@{\hspace{-0.1in}}c}
\hspace{-1.cm}\includegraphics[width=4.2cm]{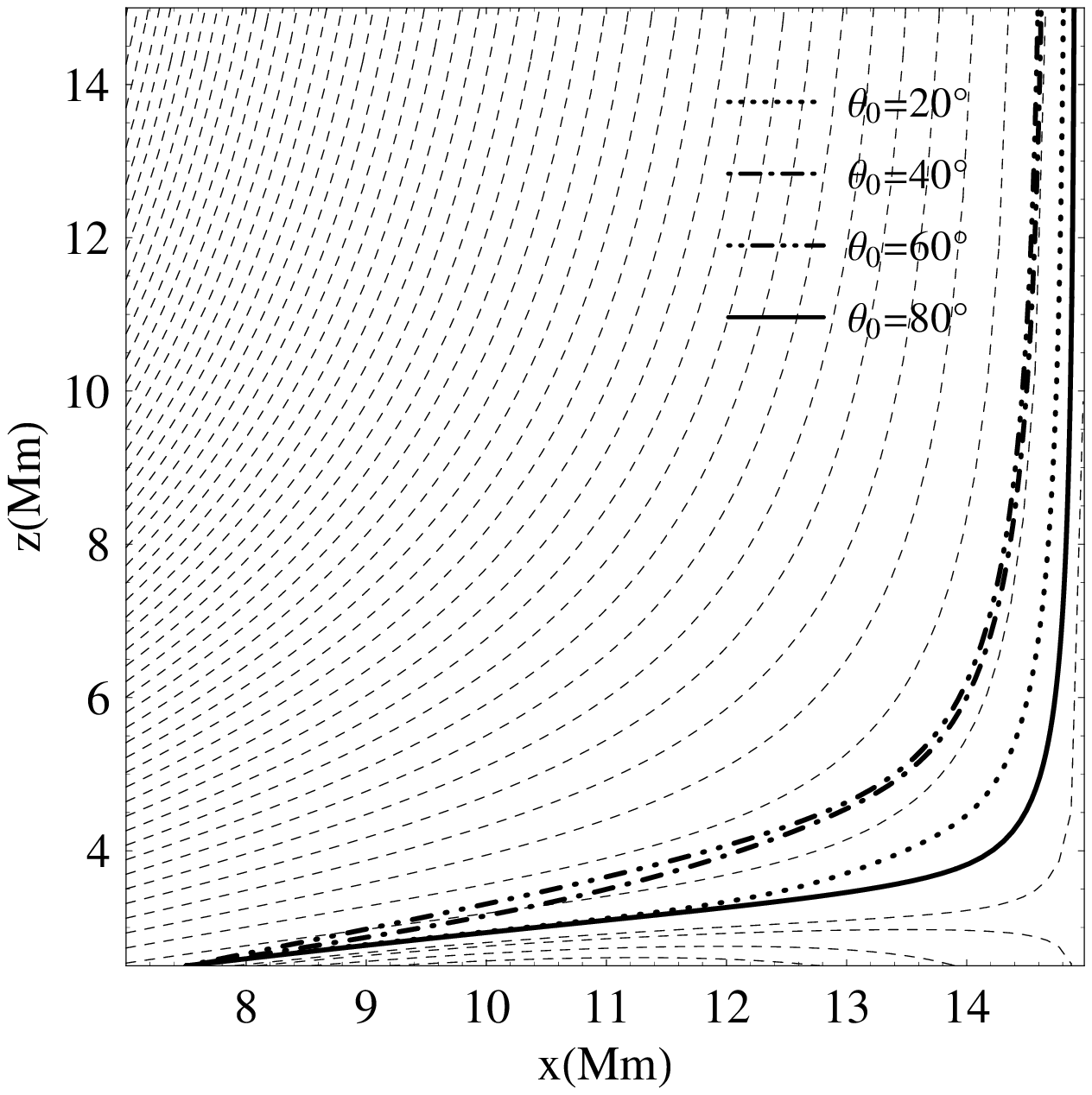}&
\hspace{-1.2cm}\includegraphics[width=4.2cm]{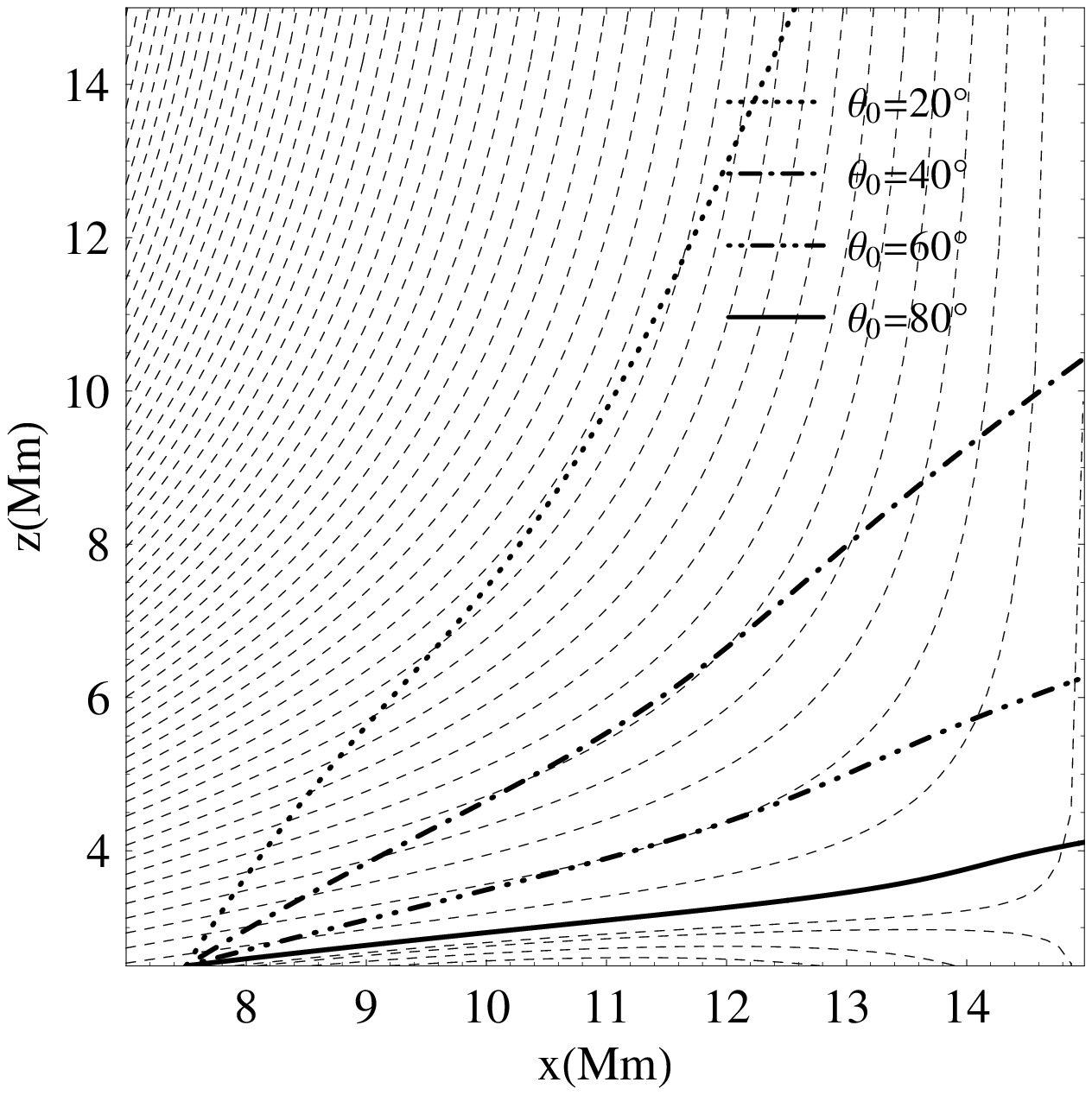}&
\hspace{-1.2cm}\includegraphics[width=4.2cm]{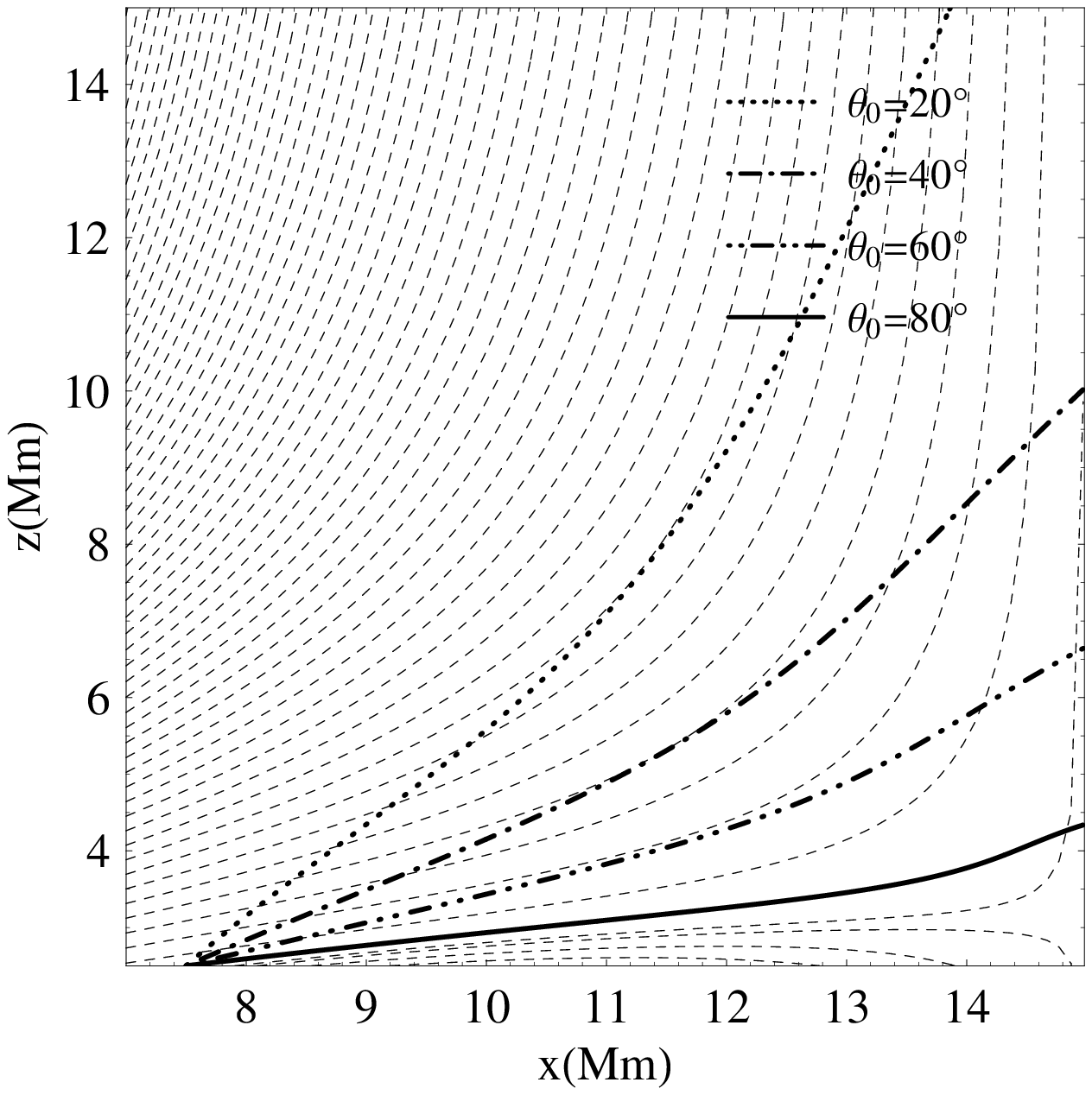}\\[-0.2cm]
\hspace{-1.cm}\includegraphics[width=4.3cm]{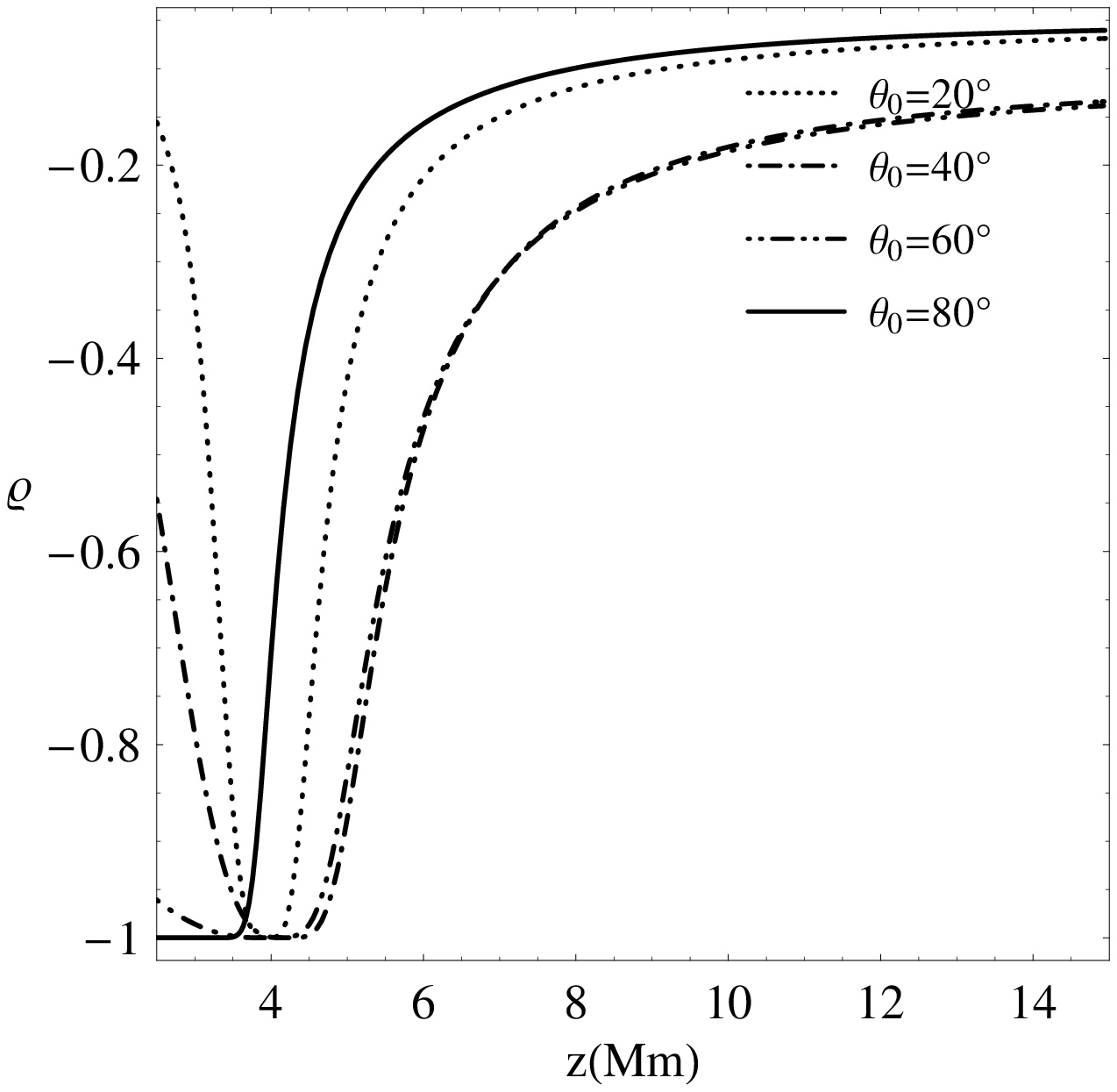}&
\hspace{-1.2cm}\includegraphics[width=4.2cm]{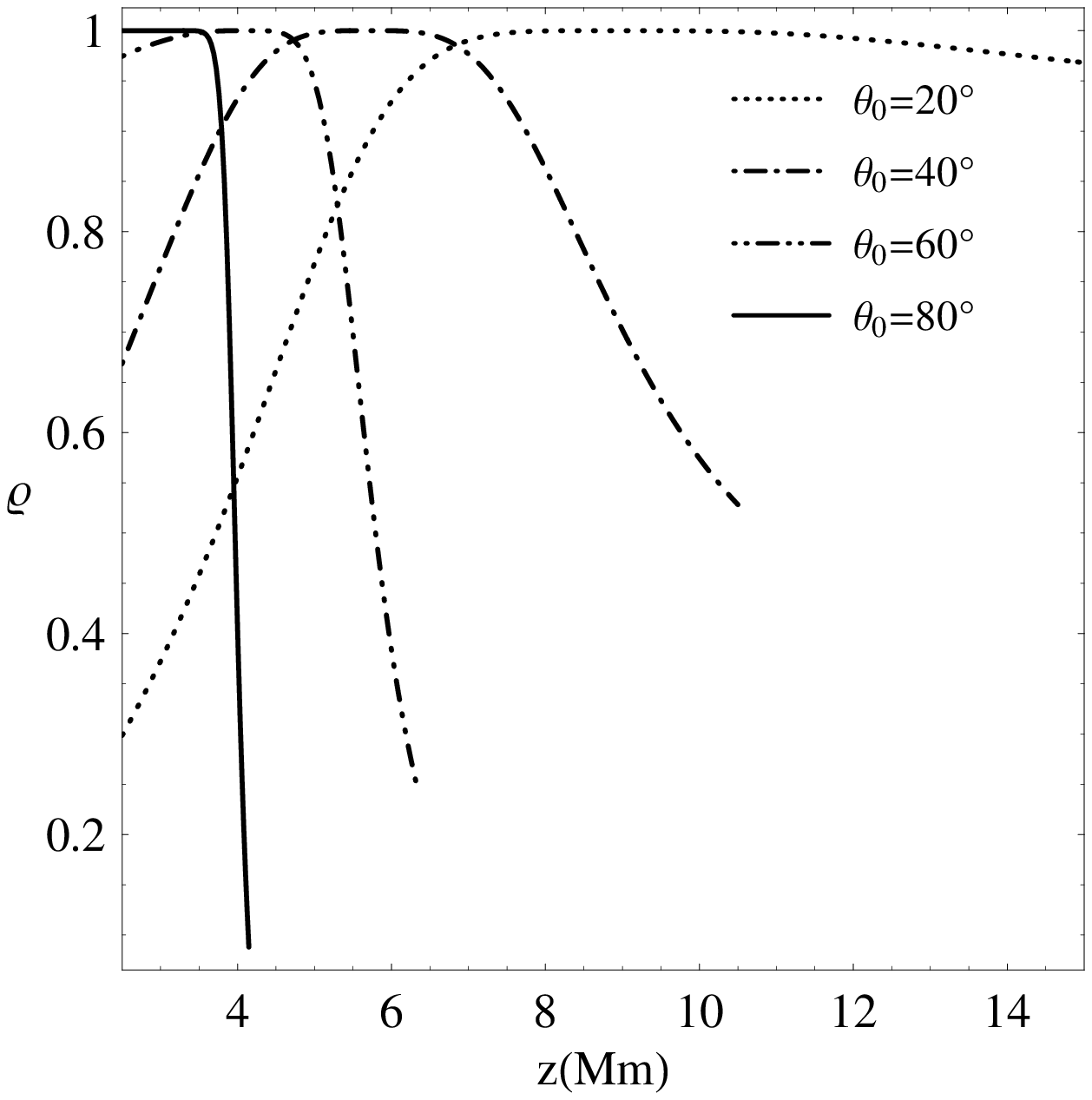}&
\hspace{-1.2cm}\includegraphics[width=4.3cm]{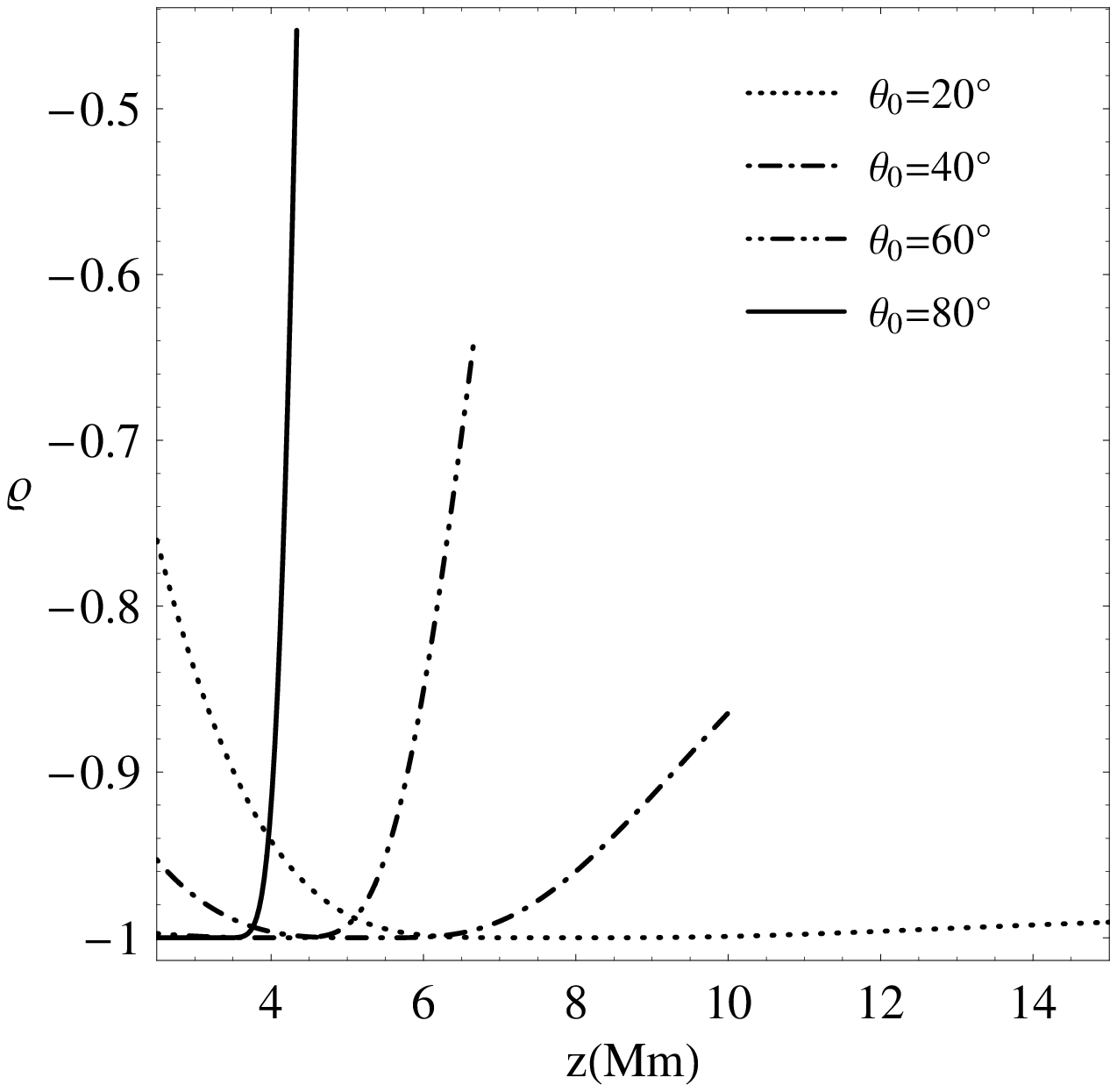}\\[-0.2cm]
\hspace{-1.cm}\includegraphics[width=6.cm]{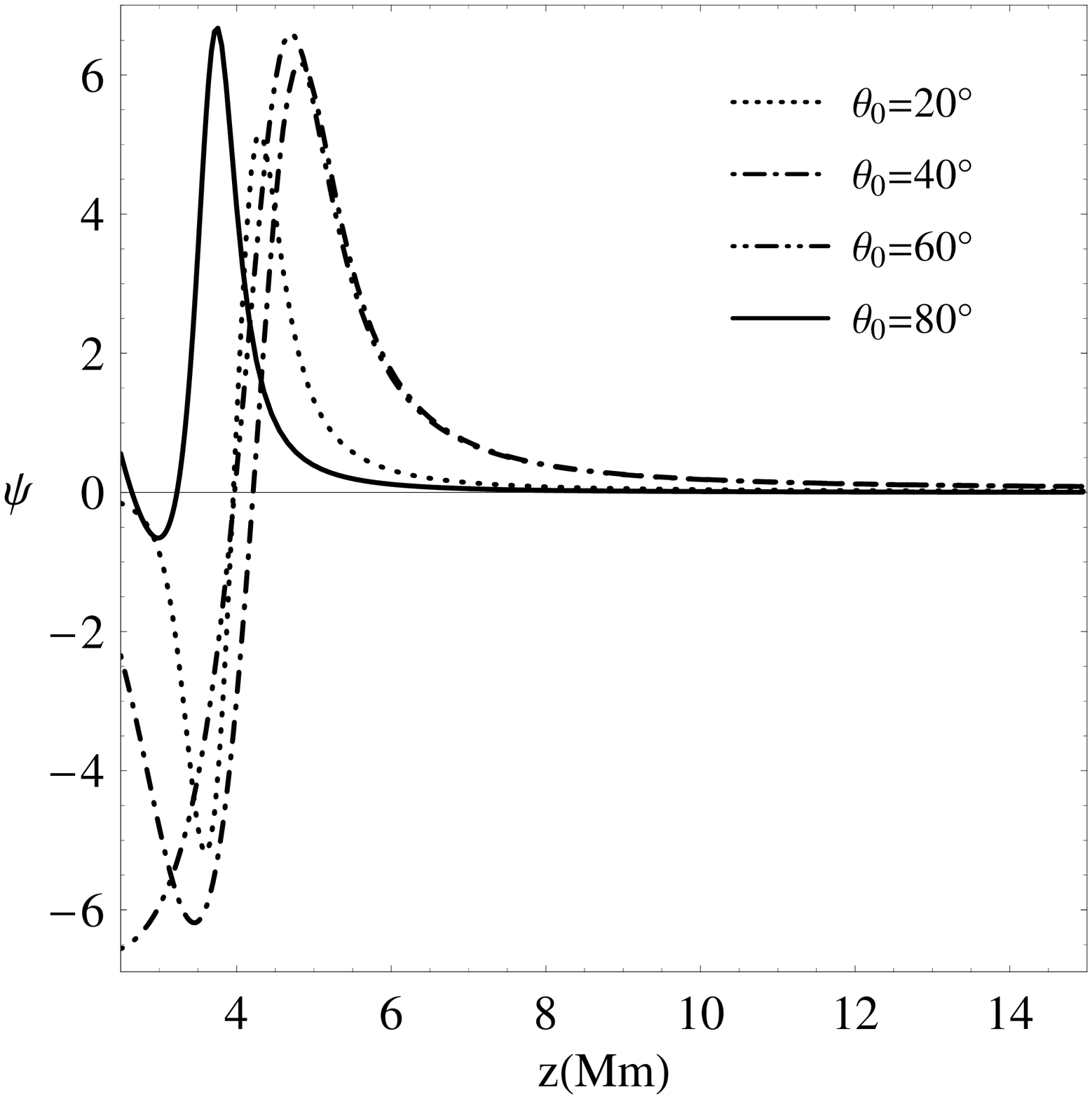}&
\hspace{-1.2cm}\includegraphics[width=6.cm]{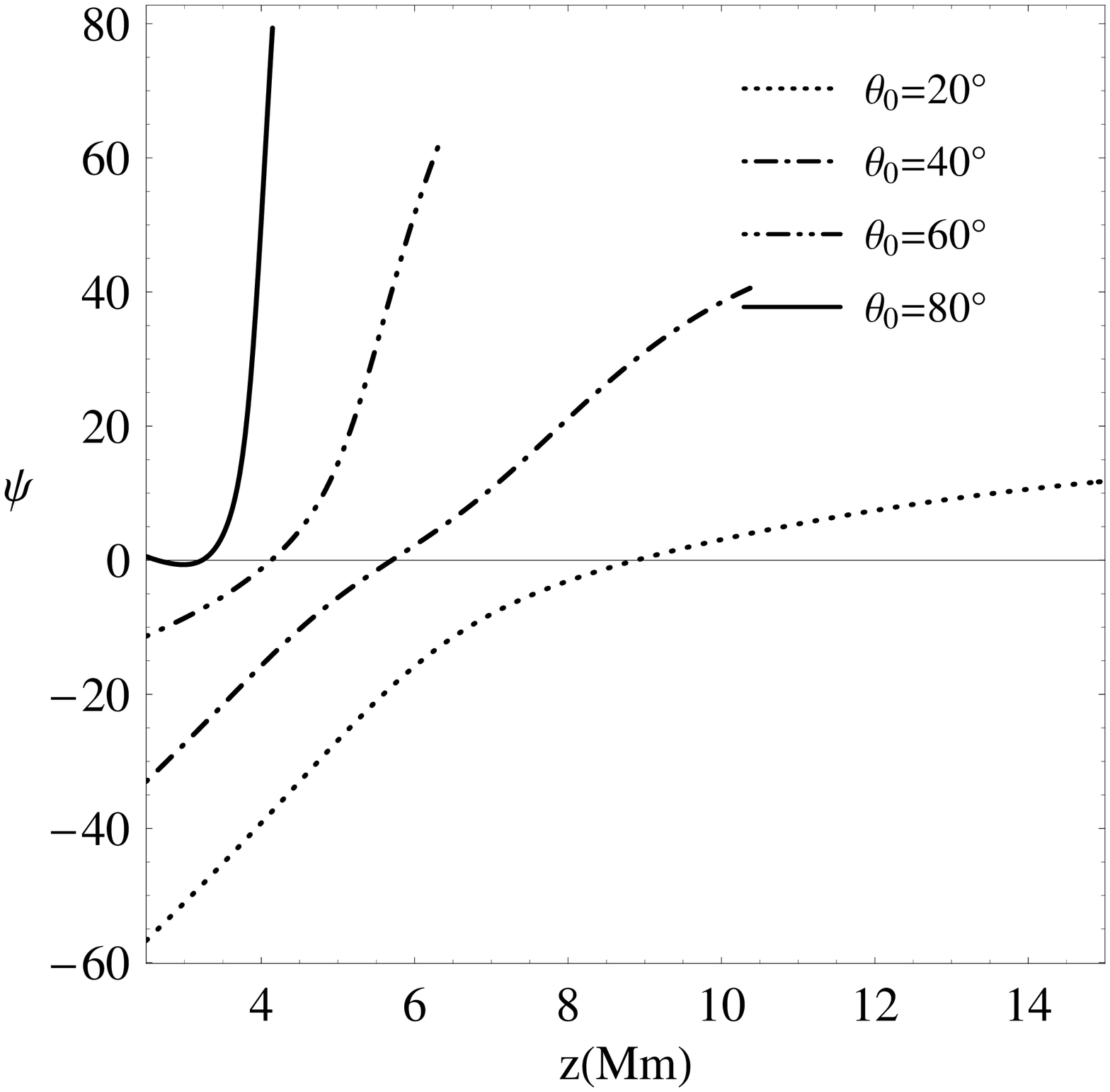}&
\hspace{-1.2cm}\includegraphics[width=6.cm]{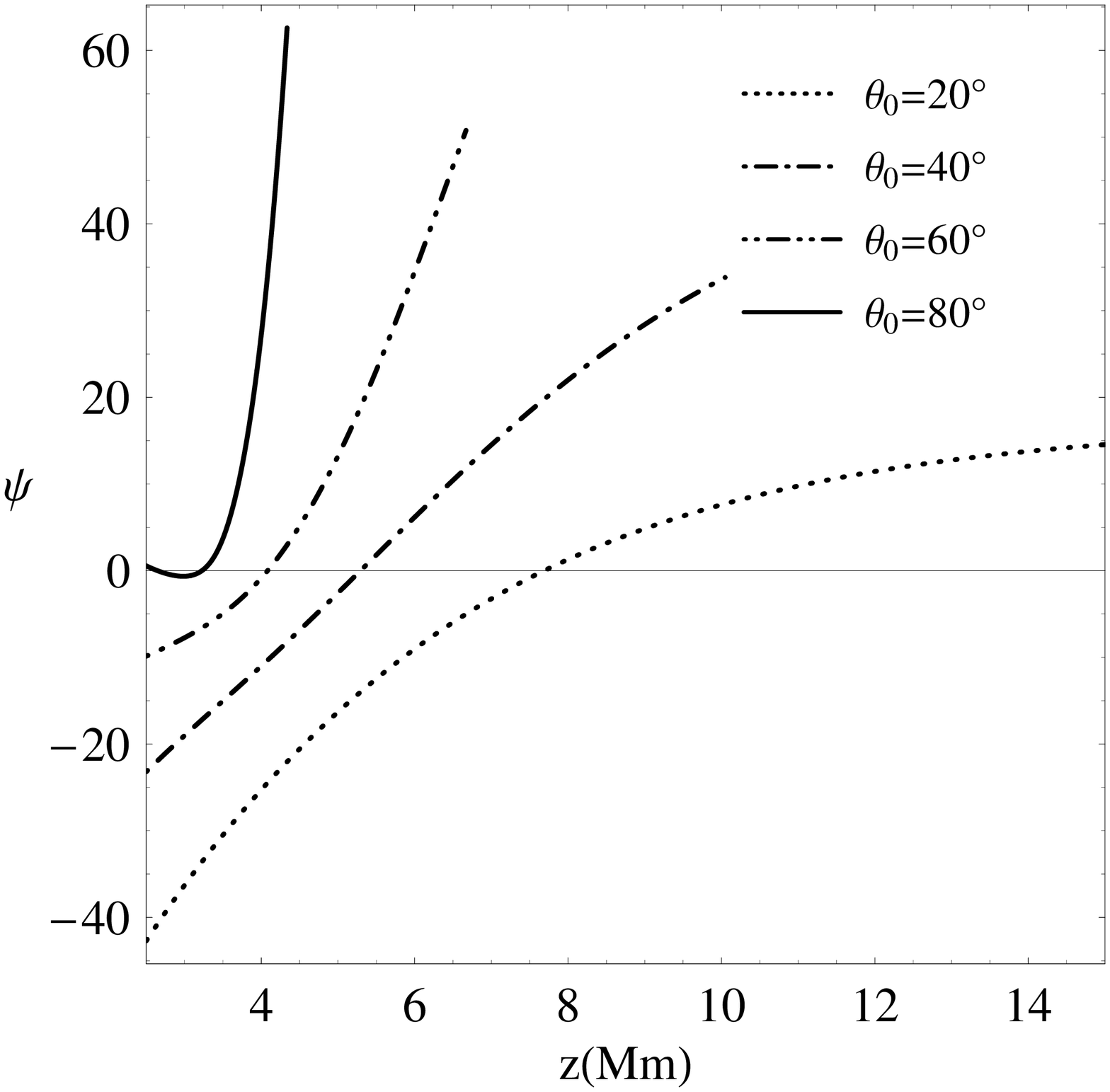}\\[-0.3cm]
\hspace{-1.2cm}\includegraphics[width=6.1cm]{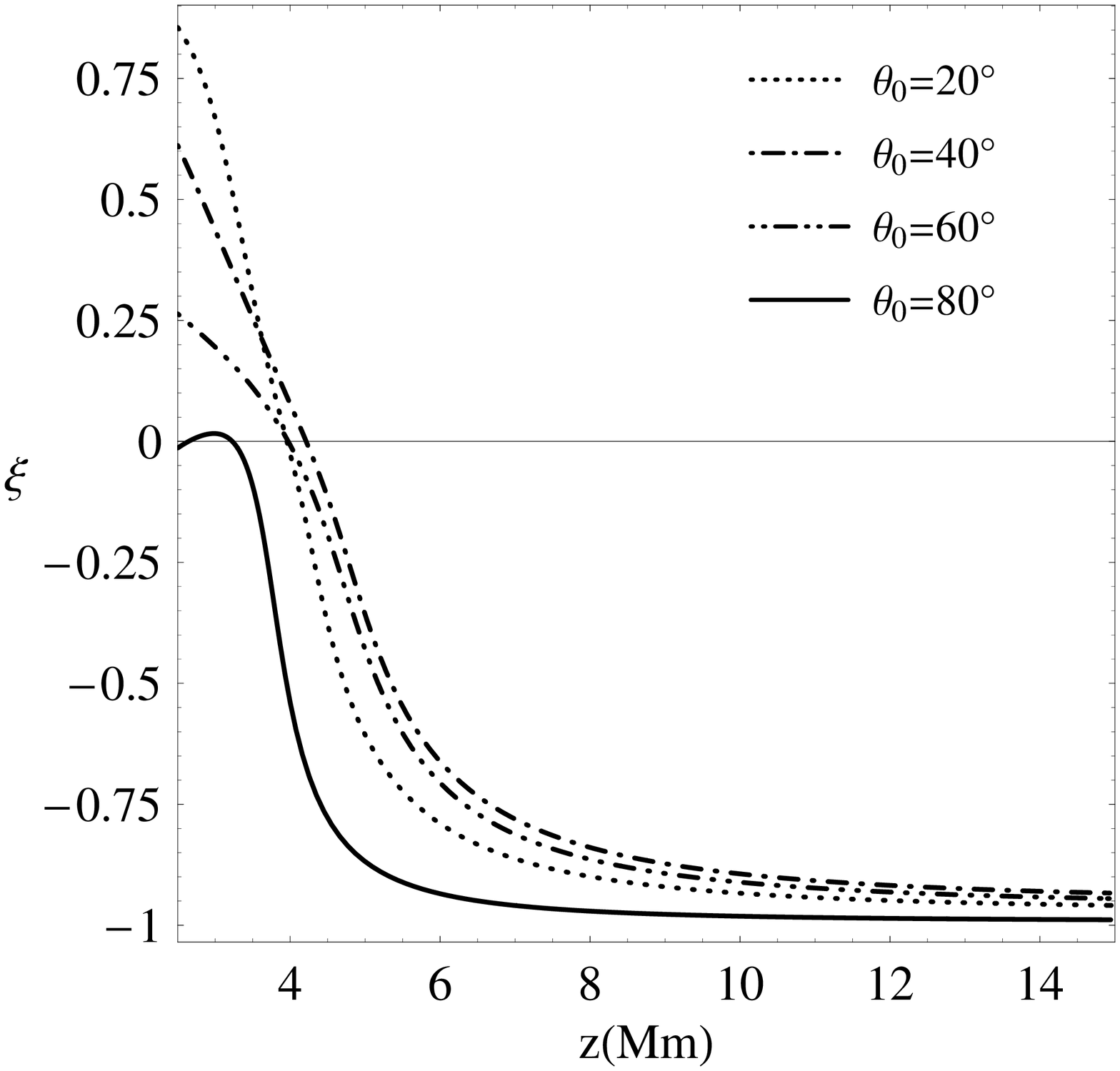}&
\hspace{-1.4cm}\includegraphics[width=6.cm]{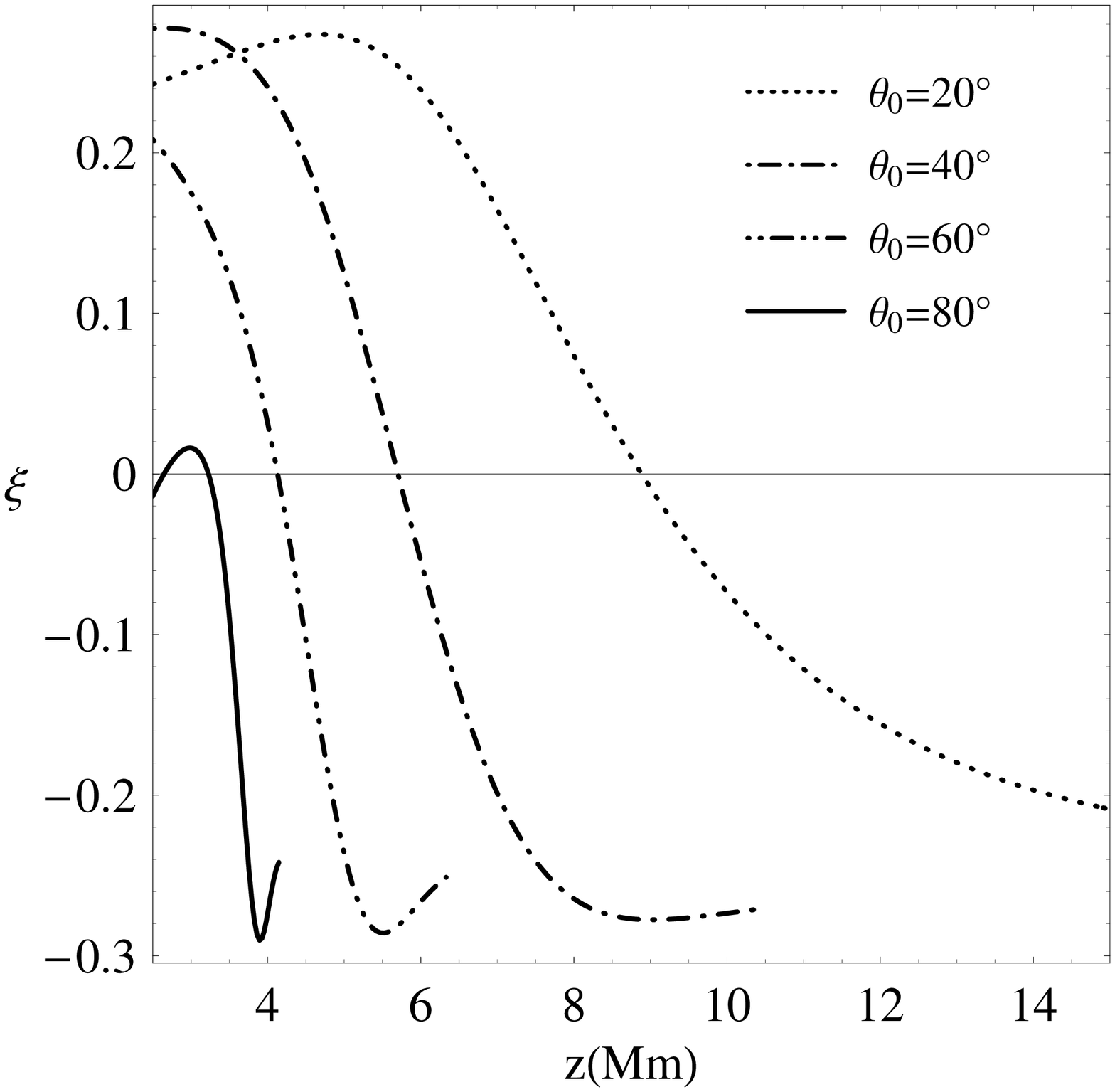}&
\hspace{-1.4cm}\includegraphics[width=6.cm]{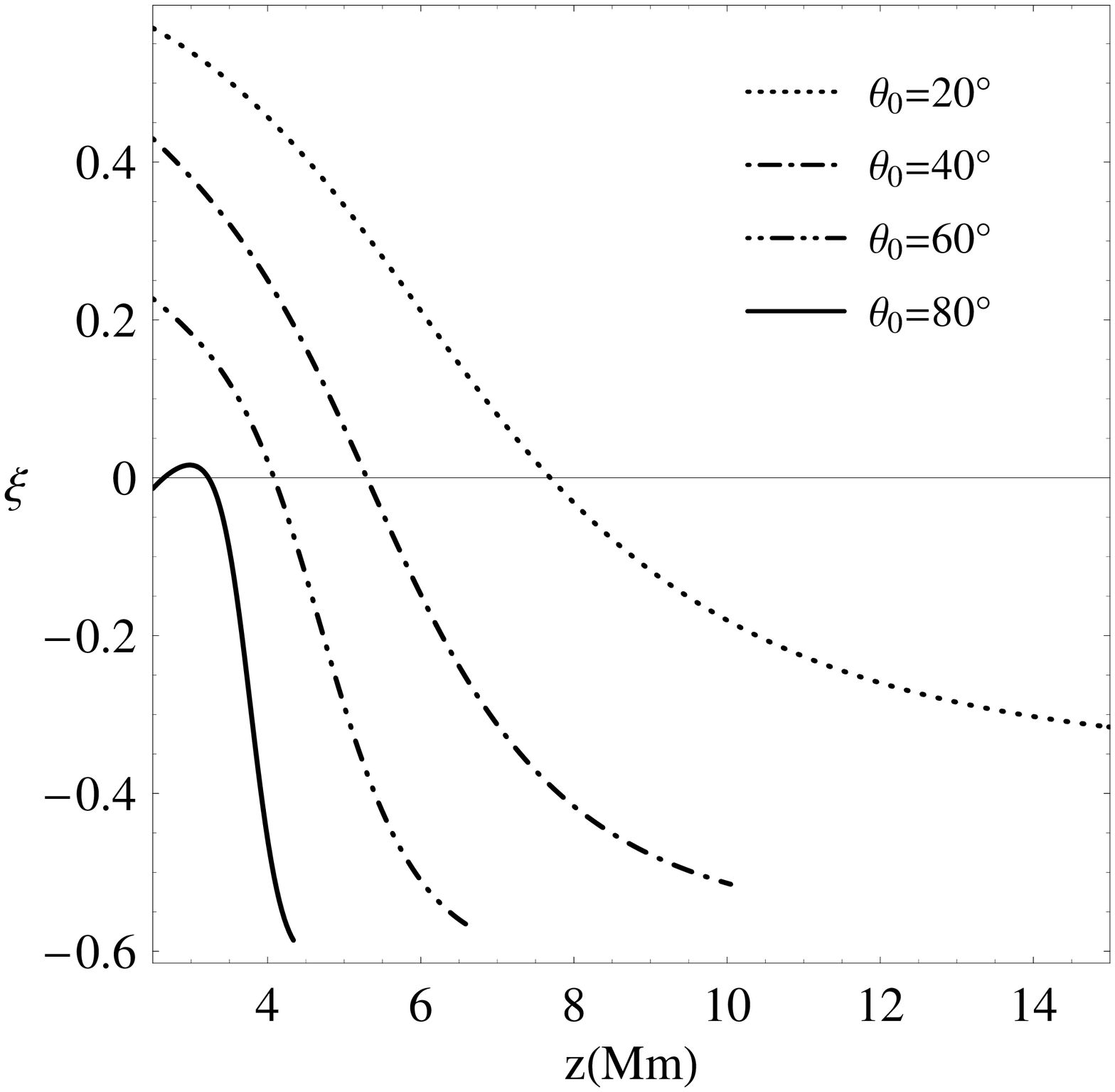}\\[-0.4cm]
\end{array}$
\end{center}
\vspace{-0.1cm} \caption{Ray paths of the waves and the spatial
variation of their basic properties. The waves are launched at the
location (x$_{0}$=7.5 Mm, z$_{0}$=2.5 Mm) in the coronal funnel
(with a \textbf{B}$_{0}$-inclination angle
$\varphi_{0}\approx79^{\circ}$) with an initial normalized wave
number k$_{0}$=0.2 and different initial angles of propagation
$\theta_{0}$. In the top panels, the dashed lines represent the
funnel field lines. The wave properties are: ($\varrho$) helicity
(degree of circular polarization), ($\psi$) angle between the
direction of the group velocity and \textbf{B}$_{0}$, ($\xi$) the
electrostatic part of the wave.} \label{RT}
\end{figure}

We consider a cold plasma made of electrons ($n_{e}$), protons
($n_{p}$) and alpha particles (He$^{2+}$, indicated by $\alpha$)
with $n_{\alpha}=0.1n_{p}$. In this case, the dispersion relation
(\ref{dispesion}) is a quadratic polynomial of degree 6, which means
that 3 modes exist and each one is represented by an oppositely
propagating ($\omega>0$ and $\omega<0$) pair of waves. These modes,
for which the dispersion curves are shown in the right panel of Fig.
\ref{VGR}, are the ion-cyclotron modes 1 (IC1) and 2 (IC2) and the
fast mode. For large k, the two IC modes reach a resonance regime at
each one of the ion-cyclotron frequencies, i.e. $\omega=\Omega_{p}$
and $\omega=\Omega_{\alpha}=\Omega_{p}/2$. The fast mode has a
cut-off frequency $\omega_{co}/\Omega_{p}=0.583$ and couples and
mode converts with the IC1 mode at the so-called cross-over
frequency $\omega_{cr}/\Omega_{p}=0.612$.
The polar plots of the group velocity for $kV_{Ap}$/$\Omega_{p}$=0.2
(Fig.~\ref{VGR}), show that the IC2 mode is mainly anisotropic and
cannot propagate perpendicularly to \textbf{B}$_{0}$. Thus the
corresponding energy mainly flows along the magnetic field lines,
with maximum group velocity at parallel propagation. On the other
hand, the IC1 mode and the fast mode propagate almost isotropically,
and consequently the energy flow is fairly isotropic.
These results are confirmed by the ray-tracing computation which
clearly shows that the IC2 mode is well guided along the field
lines, since the ray paths (direction of the group velocity) for
various initial angles of propagation $\theta_{0}$ nicely follow the
magnetic field lines (Fig.~\ref{RT}). Indeed, the maximal angular
deviation is $\psi\approx6.5^{\circ}$ at z $\approx$ 4 Mm and
$\psi\approx0^{\circ}$ (quasi-parallel propagation) in the upper
part of the funnel, where the wave is quasi-electrostatic, i.e.
$\mid\xi\mid\approx1$, and has a nearly linear polarization, i.e.
$\mid\varrho\mid\approx0$. On the contrary to that behavior, the IC1 mode and the fast mode
are unguided with a ray path having mainly a strait trajectory. The
angular deviation $\psi$ between the ray path and \textbf{B}$_{0}$
varies between $-60^{\circ}$ and $80^{\circ}$ for the IC1 mode and
$-40^{\circ}$ and $60^{\circ}$ for the fast mode. The polarization
is in general elliptical, right-handed ($\varrho>0$) for the IC1
mode and left-handed ($\varrho<0$) for the fast mode, except for
$\theta=20^{\circ}$ in the upper part of the funnel (z$\gtrsim6$ Mm)
where it is mainly circular, right-handed ($\varrho=-1$) for the IC1
mode and left-handed ($\varrho=-1$) for the fast mode.

%
%
\bibliographystyle{aa}
\bibliography{mecheri-ray}
\end{document}